\documentclass[pra,aps,groupedaddress,twocolumn,showpacs,floatfix]{revtex4}
\usepackage{amsmath,amsfonts,amssymb,graphics,graphicx,epsfig,color,times,natbib,subfigure}

\newcommand{\be}{\begin{equation}}
\newcommand{\ee}{\end{equation}}
\newcommand{\ba}{\begin{eqnarray}}
\newcommand{\ea}{\end{eqnarray}}
\newcommand{\ban}{\begin{eqnarray*}}
\newcommand{\ean}{\end{eqnarray*}}

\newcommand{\ket}[1]{\mbox{$ | #1 \rangle $}}
\newcommand{\bra}[1]{\mbox{$ \langle #1 | $}}

\newcommand{\si}{\sigma}

\newcommand{\one}{\leavevmode\hbox{\small1\normalsize\kern-.33em1}}

\begin{document}

\title{Testing a Bell inequality in multi-pair scenarios}
\author{Jean-Daniel Bancal$^1$, Cyril Branciard$^1$, Nicolas Brunner$^1$, Nicolas Gisin$^1$, Sandu Popescu$^{2,3}$, and Christoph Simon$^1$}
\address{$^1$Group of Applied Physics, University of Geneva, 20 rue de l'Ecole-de-M\'edecine, CH-1211 Geneva 4, Switzerland\\
$^2$H.H. Wills Physics Laboratory, University of Bristol, Tyndall Avenue, Bristol BS8~1TL, U.K.\\
$^3$Hewlett-Packard Laboratories, Stoke Gifford, Bristol BS12~6QZ, U.K.}
\date{\today}

\begin{abstract}
To date, most efforts to demonstrate quantum nonlocality have concentrated on systems of two (or very few) particles. It is however difficult in many experiments to address individual particles, making it hard to highlight the presence of nonlocality. We show how a natural setup with no access to individual particles allows one to violate the CHSH inequality with many pairs, including in our analysis effects of noise and losses. We discuss the case of distinguishable and indistinguishable particles. Finally, a comparison of these two situations provides new insight into the complex relation between entanglement and nonlocality.
\end{abstract}
\maketitle


\section{Introduction}

Entanglement is the resource that allows one to establish quantum
non-local correlations \cite{Bell64}. These correlations have been
the center of a wide interest,  because of their fascinating
nature, and of their impressive power for processing information.
Experimentally, quantum non-locality is demonstrated in so-called
Bell experiments, which have to date all confirmed the quantum
predictions \cite{aspect99}.


Most theoretical works on Bell experiments and Bell inequalities
have focused on the case where the source emits a single entangled
pair of particles at a time. Indeed, this is the simplest situation to study.
From the experimental point of view, most experiments have been
designed in order to match this theoretical model. For example, in
photonic experiments, the source, usually based on parametric down
conversion (PDC), is set in the weak regime; i.e. when the
source emits something, it is most likely a single pair of
entangled photons.

However, there are experimental situations, such as in many-body
systems, where producing single entangled pairs is rather
difficult. For instance, in Ref. \cite{anderlini} many entangled
pairs ($\simeq 10^4$) of ultracold atoms have been created but
cannot be addressed individually. So, while entanglement has
definitely been created in this system, one still lacks an
efficient method for demonstrating its quantum non-locality
through the violation of some Bell inequality. The goal of the
present paper is to discuss techniques for testing Bell
inequalities in such multi-pair scenarios, where the particles on
Alice's and Bob's side cannot be individually addressed, and must
therefore be measured globally (see Fig. 1). What we mean here by 
global measurements is that each particle is submitted to the
same measurement. Note that the case of more general measurements
(collective measurements on all particles) has been considered in
Ref. \cite{doherty06}.

Basically one should distinguish two cases: independent pairs and
indistinguishable pairs. In the first case, the pairs are created
independently, but cannot be addressed individually; therefore
they must be measured globally (both on Alice's and on Bob's
side). During this global measurement, the classical information
about the pairing is lost: there is no way to tell which particle
is entangled with which. The corresponding loss of entanglement
has been derived in Ref. \cite{eisert00}. In the second case, the
pairs are indistinguishable; so in some sense the information
about the pairing is here lost in a coherent way.

Reid \textit{et al.} \cite{reid02} have considered the case of
indistinguishable pairs (with global measurement) in optics. More
specifically, these authors, extending on a previous work of Drummond
\cite{drummond83}, showed how Bell inequalities can be tested (and
violated) when many pairs are created via PDC. In this case the
pairs are indistinguishable because of the process of stimulated
emission. In Ref. \cite{inept}, Jones \textit{et al.} have
considered a related scenario; there entangled pairs are delivered
via an inept delivery service, but at the end only a single pair
is measured. Also considering multi-particle entanglement in such
scenario is an interesting problem~: see for example
\cite{toth07,eisenberg04}.

In this paper we will study the violation of Bell inequalities in
a general multi-pair scenario. We start by treating the case of
independent pairs (Section \ref{indep}). We argue that the resistance to
noise is here the relevant measure of non-locality, evaluating it.
The consequences of particle losses are also investigated. Next we move
to the case of indistinguishable pairs (Section \ref{indist}), after a brief
review of the results of Ref. \cite{reid02}, we present an
analysis of the influence of noise and losses in this case. In Section \ref{comparison}, we
compare the entanglement and non-locality in both cases. This leads
us to a surprising result: while the state of indistinguishable
pairs contains more entanglement than the state of independent
pairs (after the classical mixing), the latest appears to be more
non-local. In other words, the incoherent loss of information
provides more non-locality, but less entanglement, than the
coherent loss of information (indistinguishable pairs). This
provides a novel example (here in the case of multi-pairs) of the
complex relation between entanglement and non-locality. Finally we
provide some experimental perspectives (Section \ref{experiments}) and conclusions.

\section{Independent pairs}\label{indep}

We consider a source emitting $M$ entangled pairs, each of them being in the same entangled two-qubit state $\rho$. Thus the global state is

\ba\label{eq:etat} \boldsymbol{\rho}_M =  \rho^{\otimes M} = \underbrace{\rho  \otimes \rho \otimes \dots \otimes \rho}_{M\text{ times}} .\ea

\begin{figure}
\includegraphics[width=8.5cm]{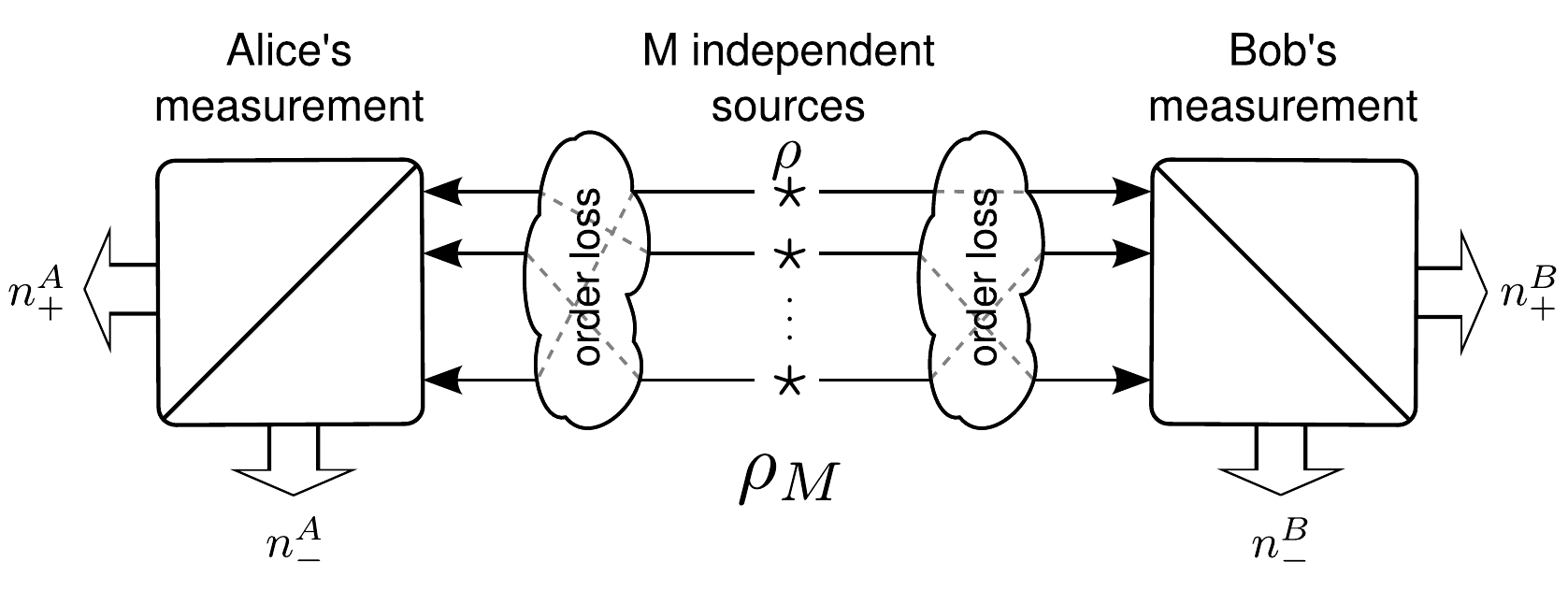}
\caption{Setup : a source produces $M$ independent pairs (or equivalently M independant sources each produce a pair), the pairing between Alice's and Bob's particles is lost during their transmission, and each party measures all their incoming particles in the same basis. The total number $n_{+(-)}$ of particles detected in the $+(-)$ outcome is tallied on both sides.}
\label{fig:setup}
\end{figure}

Each pair being independent, Alice and Bob receive $M$ uncorrelated particles. Since Alice and Bob are unable to address single particles in their ensemble, they perform a global measurement on their $M$ particles, i.e. all $M$ particles are measured in the same basis (we shall consider here only von Neumann measurements), or equivalently in the same direction on the Bloch sphere. After the measurement apparatus, two detectors count the number of particles, $n_+$ and $n_-$ in each output mode (see figure \ref{fig:setup}). If the detectors are perfectly efficient ($\eta=1$), one has $M = n_+ + n_-$.

\subsection{Testing the CH inequality}

Our goal is to test a Bell inequality. Here we shall focus on the simplest Bell inequality, the CHSH inequality \cite{clauser69}, which involves two inputs on Alice and Bob's sides $A_1, A_2$ and $B_1, B_2$, and two outputs, $\alpha,\beta\in\{+,-\}$. For convenience we write it under the CH form \cite{clauser74},

\be\label{CH}
\begin{split}
CH=-P_+^A(A_1)-P_+^B(B_1)+P_{++}(A_1,B_1)\\ +P_{++}(A_1,B_2)+P_{++}(A_2,B_1)-P_{++}(A_2,B_2)\leq0 
\end{split}
\ee where $P_{++}(A_i,B_j)$ is the probability for both Alice and Bob to output "+" when performing measurements $A_i$ and $B_j$ respectively. Recall that under the hypothesis of no-signaling both CH and CHSH inequalities are equivalent \cite{collins04}. Now, in order to test inequality \eqref{CH}, Alice and Bob must transform their data, basically $n_+$ and $n_-$, into a binary result, ``$+$'' or ``$-$''. A natural way of doing it is by invoking a voting procedure, for instance:
\be\label{eq:voting}
\begin{split}
1.&\textrm{Majority voting: if $n_+ \geq n_-$ } \rightarrow  \textrm{``} + \textrm{'', otherwise ``} -\textrm{''} \\
2.&\textrm{Unanimous voting: if $n_+ = M$ } \rightarrow  \textrm{``} + \textrm{'', otherwise ``} - \textrm{''}
\end{split}
\ee
or any intermediate possibility, for instance 2/3 or 3/4 majority. For each voting method and given $M$ corresponds a threshold $N=\lceil M/2\rceil,\lceil 2M/3\rceil,...,M$ such that the outcome is $+$ iff $n_+ \geq N$.

At this point the two relevant questions are : first, is it possible to violate the CH inequality with any of these voting procedures? Second, if yes, which strategy yields the largest violation? To address these questions one must compute the joint and marginal probabilities entering the CH inequality for each procedure.

\subsection{Pure states}

Let us first consider the pure entangled states
\be\label{eq:psi}
\ket{\psi}=\cos{\theta}\ket{00} + \sin{\theta}\ket{11}
\ee
so that in equation \eqref{eq:etat}, $\rho=\ket{\psi}\bra{\psi}$. We will also write $\ket{\Psi_M}=\ket{\psi}^{\otimes M}$. For detectors with a perfect efficiency $\eta =1$, all $M$ particles are detected both on Alice's and on Bob's side. The marginal and joint probabilities entering the CH inequality for a vote with given threshold $N$ are

\ba\label{eq:marginal}
P_+(A)=\sum_{n_+=N}^M \binom{M}{n_+}p_+(A)^{n_+}p_-(A)^{M-n_+} \\
P_{++}(A,B)=M!\sum_{n_{\alpha\beta}\in\Xi}\prod_{\alpha,\beta} \frac{p_{\alpha\beta}(A,B)^{n_{\alpha\beta}}}{ n_{\alpha\beta}!}
\ea where $p_{++}(A_i,B_j)=tr(A_i \otimes B_j \ket{\psi}\bra{\psi})$ and $p_+(A_i/B_j)=tr(A_i \otimes \one/\one \otimes B_j \ket{\psi}\bra{\psi})$ are the quantum joint and marginal probabilities for a single pair. Alice and Bob's outputs are noted $\alpha,\beta\in\{+,-\}$, $n_{\alpha\beta}$ is the number of pairs which gave detections $\alpha$ and $\beta$, and $\Xi=\{n_{\alpha\beta}\in \mathbb{N}_+|\sum n_{\alpha\beta}=M, n_+^A=n_{++}+n_{+-}\geq N, n_+^B=n_{++}+n_{-+}\geq N\}$ is the set of all events yielding the result ``$++$'' after voting.

Next one can choose the state $\ket{\psi}$ and the measured settings. For the maximally entangled state of two qubits ($\theta=\pi/4 $) one may choose the standard optimal (for the case $M=1$) settings for the CH inequality, i.e. $A_0 = \si_z$, $A_1 = \si_x$, $B_0 = \frac{\si_x+\si_z}{\sqrt2}$, $B_1 = \frac{-\si_x+\si_z}{\sqrt2}$. Doing so with majority voting ($N=\lceil M/2 \rceil$), the CH inequality can be violated for any value of $M$; the maximal amount of violation is numerically found to decrease with the number of emitted pairs as $M^{-1}$.

\begin{figure}
\subfigure[]{\label{fig:CH}\includegraphics{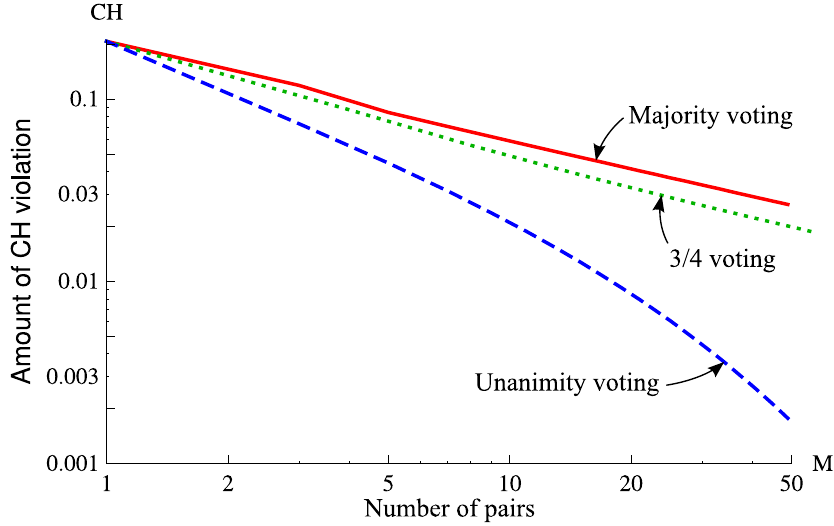}}
\subfigure[]{\label{fig:1mw}\includegraphics{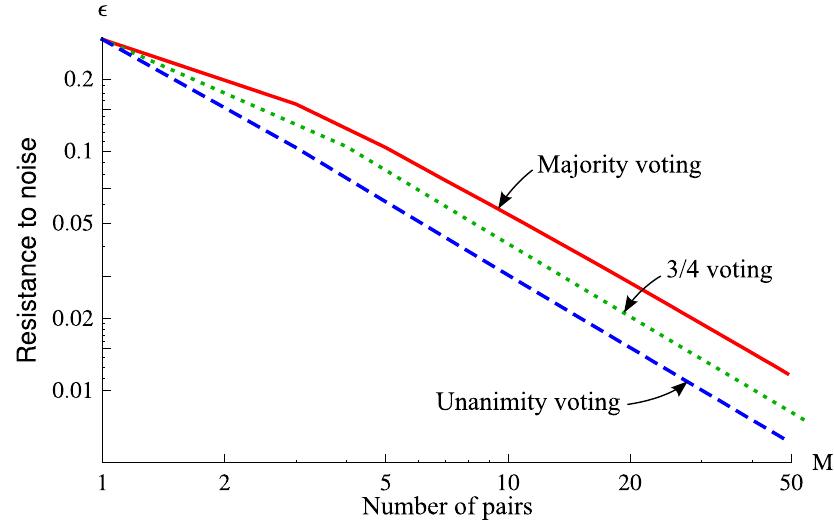}}
\caption{(Color online) CH violation and resistance to noise for a source producing $M$ independent pairs. The states and settings used are discussed in the text.\\
(a) Maximal CH values for various thresholds~: majority voting (full red line), 3/4 voting (dotted green line) and unanimity (dashed blue line). The decrease is as $M^{-1/2}$ for the first two, and exponential for the last one. The highest violation is thus reached using a majority vote.\\
(b) Resistance to noise for the different thresholds (same colors). All curves decrease as $M^{-1}$. The most resistant violation is that achieved by using majority voting.}
\end{figure}

Remarkably, a higher violation is found for different measurement settings, given by
\be\label{eq:settings}
\begin{split}
A_0&=\si_z\\
A_1&=\sin2\alpha\si_x+\cos2\alpha\si_z\\
B_0&=\sin\alpha\si_x+\cos\alpha\si_z\\
B_1&=-\sin\alpha\si_x+\cos\alpha\si_z.
\end{split}
\ee
With $\alpha\sim \frac{\pi}{2\sqrt{2}}M^{-1/2}$, those settings are numerically found to be optimal. In this case the decrease of $CH$ is only $M^{-1/2}$ (see figure \ref{fig:CH}). The state leading to the largest violation is always the maximally entangled one ($\theta=\pi/4$) for majority voting.

The one-parameter planar settings \eqref{eq:settings} were already used by several authors \cite{reid02,drummond83}, for example in Bell experiments using a state we shall look at later \eqref{eq:phi} with the unanimous vote for any $M$.

Performing numerical optimizations, we also found that a violation can be obtained for any voting strategy with any number of emitted pairs $M$ (see figure \ref{fig:CH}). We optimized the state ($\theta$) and the four measurement settings, each time finding optimal settings of the form \eqref{eq:settings}. For the unanimous vote, for instance, the optimal state is less and less entangled as the number of emitted pairs $M$ increases, as described in \cite{brunner08}, and the violation decreases exponentially with $M$. Thus for pure entangled 2-qubit states, the largest amount of violation is obtained with majority voting.

\subsection{Resistance to noise}\label{Wernerstates}
We now compute the resistance to noise that these violations could bear, which is the relevant measure of non-locality considering experimental perspectives --- the amount of violation being basically just a number, without much significance in the present case as we shall see.

In a practical EPR experiment, imperfect detectors, noisy sources or disturbing channels introduce noise in the measurement results. To first order, this noise can be modeled at the level of the source, supposing that the produced pairs are not in the pure state $\ket{\psi}\bra{\psi}$, but instead in a Werner state of the form
\be \label{eq:werner} \rho = w\ket{\psi}\bra{\psi}+(1-w)\frac{\one}4 .\ee
The resistance to noise of a given violation is then defined by the maximal amount $\epsilon=1-w$ of white noise that can be added to the pure state $\ket{\psi}\bra{\psi}$ such that the resulting state $\rho$ still violates a Bell inequality (CH here).

Considering now that the sources of figure \ref{fig:setup} produce the state \eqref{eq:werner}, we look for the largest value of $\epsilon$ which still gives a positive value of $CH$, using settings of the form \eqref{eq:settings} and optimizing on the state ($\theta$). For all voting strategies we find a resistance to noise decreasing like $\epsilon\sim M^{-1}$, the majority voting being still the best choice (see Fig. \ref{fig:1mw}). Unlike when maximizing $CH$ in the absence of noise, here the optimal state is always the maximally entangled one ($\theta=\pi/4$), even for intermediary voting strategies, for which the $CH$ violation with this state decreases exponentially with $M$. This shows that appropriate figures of merit need to be used when examining practical situations.

These results are encouraging, but just as detectors might not be perfect, maybe the source cannot guarantee an exact number of pairs $M$, as needed here. To show that these violations are relatively robust towards this issue, we now look at the case of sources producing a number of entangled pairs which follow a poissonian distribution.

\subsection{Poisson sources}

A poissonian source produces a state $\boldsymbol{\rho}_M$ of $M$ pairs with a poissonian probability
\be p(M)=e^{-\mu}\frac{\mu^M}{M!}\ee
where $\mu$ is the mean number of photon pairs. With such a source, a different number of pairs is created every time. So for a chosen voting assignment \eqref{eq:voting} the threshold $N$ varies with the total number of photons detected $M=n_+ + n_-$ (we still consider perfect detectors), according to each realization.

\begin{figure}
\subfigure[]{\label{fig:PoissonCH}\includegraphics{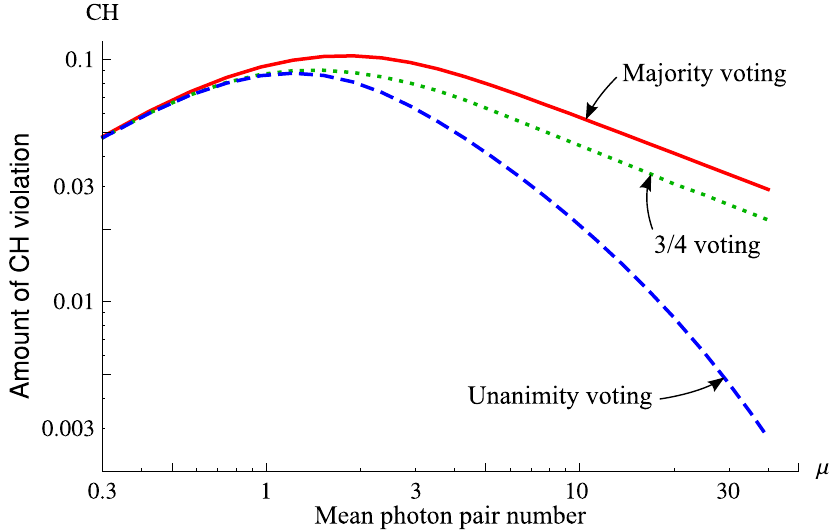}}
\subfigure[]{\label{fig:Poisson1mw}\includegraphics{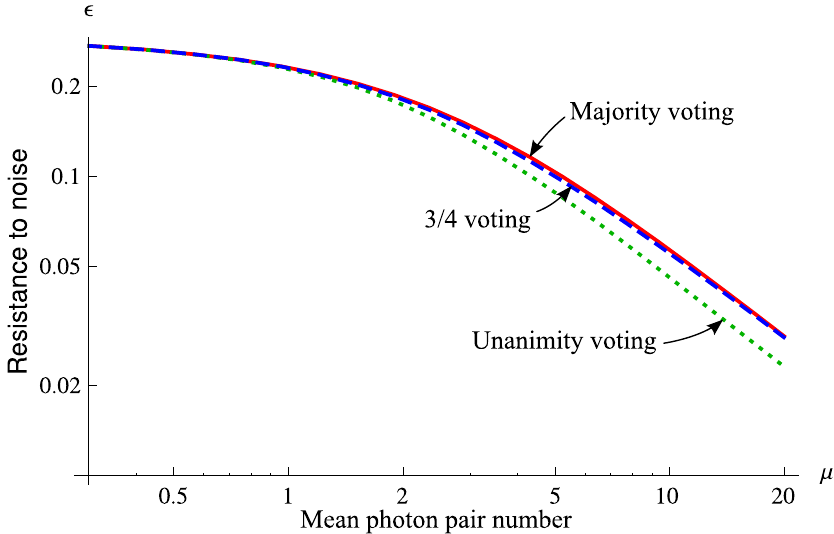}}
\caption{(Color online) Maximal Bell violation and resistance to noise with a poissonian source of independent pairs. The red lines represents the majority voting, the green dotted lines the 3/4 voting and the blue dashed lines the unanimity voting. Settings are chose in the form \eqref{eq:settings}, optimal states are discussed in the text.\\
(a) For a large mean photon number, the decrease of $CH$ goes like $\mu^{-1/2}$ for the majority and 3/4 vote, just like for the fixed $M$ case. Concerning the unanimity vote, $CH$ decreases faster than a polynomial, but slower than an exponential.\\
(b) Resistance to noise is very similar for all strategies, decreasing as $\mu^{-1}$ just like with as source of fixed pairs number.}
\label{fig:Poisson}
\end{figure}

Using settings of the form \eqref{eq:settings}, we optimized numerically $\alpha$ and the state ($\theta$) for several votes, in a situation where the source is poissonian. Doing so in order to get the largest $CH$ violation and the highest resistance to noise, we obtained results very similar to that of the fixed $M$ case, verifying in particular a decrease of $CH$ as $\mu^{-1/2}$ for the majority vote and of the resistance to noise as $\mu^{-1}$ (see fig. \ref{fig:Poisson}). Similarly, the states yielding the largest $CH$ values are the maximally entangled one for the majority vote, and partially entangled ones for the two other votes. A difference, however, is that $CH$ is found to decrease slower than exponentially for the unanimity vote.

Note also that since it is possible to find $CH>0$, and the probability to get a $+$ result vanishes for $\mu\to0$, there exist an optimal $\mu=1.2\sim1.8$ yielding a maximum CH violation. But this feature is not found in the resistance to noise.


\subsection{Inefficient detectors}

We now consider detectors with finite efficiency $\eta<1$, and look in what circumstance a Bell violation can still be observed in a multipair scheme with such detectors. $\eta$ is to be understood here as the probability for a particle to be detected.

\begin{figure}
\includegraphics{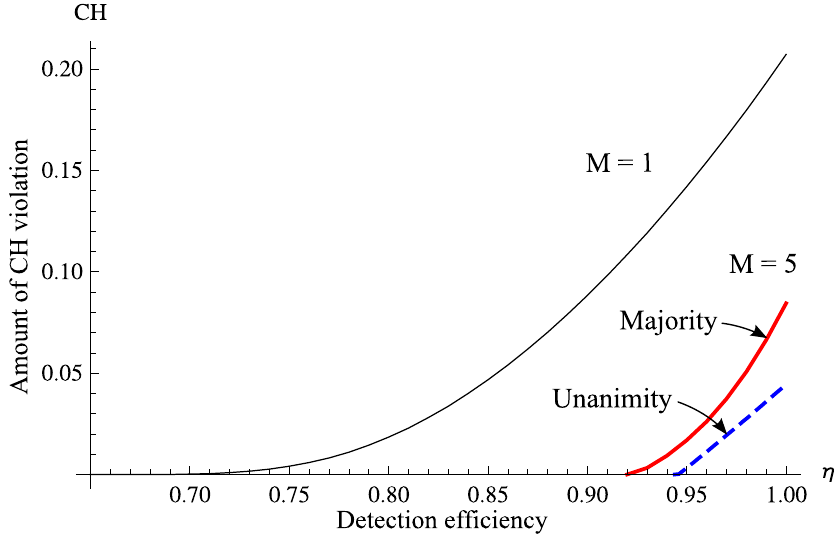}
\caption{(Color online) $CH$ violation with inefficient detectors, as a function of the probability for a photon to be detected. The upper thin curve shows the traditional case $M=1$ with known critical efficiency $\eta=2/3$ \cite{eberhard93}. The two other curves are for $M=5$ pairs with majority (full red line) and unanimity voting (dashed blue line).}
\label{fig:ineff}
\end{figure}

In general, in presence of detectors inefficiencies (or particle losses) the total number of particles detected by Alice and Bob are different ($n_+^A+n_-^A\neq n_+^B+n_-^B$). Thus, for a given voting strategy, the thresholds $N$ applied by Alice and Bob might be different for the same event, since it depends on the total number of photons detected by each party. Testing a Bell inequality in this situation without appealing to post-selection introduces no detection loophole, but it is not a surprise to find that high efficiencies are needed in order to find a Bell violation in such circumstances. Figure \ref{fig:ineff} shows the maximal CH violations obtained (optimizing on states and settings) as a function of the detection efficiency for $M=1$ and $M=5$ with majority and unanimity voting. The required detector efficiency increases with the number of pairs, leaving no chance to find a violation at high $M$ with $\eta<<1$.

One way to deal with detector inefficiencies consists in post-selecting events in which exactly $M$ photons are detected on both sides. This way cases in which particles were not detected are neglected and the Bell violation is recovered independently of the losses. (If detectors also have dark counts, noise will appear in the statistics, which can be treated with Werner states as presented in the 'resistance to noise' section.) This approach is however not perfect as it is subject to the so-called detection loophole \cite{Pearle70} : there exist local models, exploiting detectors inefficiencies, that can violate a Bell inequality \cite{gisin99}. But also, one needs to know exactly the number of pairs $M$ created, before measuring them. This last condition might not be guaranteed, for example with poissonian sources where the knowledge of $M$ is often inferred from the number of detected particles.

To estimate the impact of losses, we consider the case in which exactly 1 of the $M$ photons flying to Alice, and 1 going to Bob, are not detected. As the number of created pairs increases, this is a situation that must happen frequently even with very efficient detectors. Using the majority and unanimity vote in this situation we numerically verified that the $CH$ inequality couldn't be violated, at least for $M\leq50$.

A way to understand this result is by noting that the sets of events yielding results + and - are separated by only 1 photon number. Thus, removing one photon mixes the two sets. It should thus be advantageous to separate these two cases such that for instance $n_+ \geq N \to +$, $n_- \geq N \to -$, $M-N<n_+<N \to \varnothing$. Using this particular post-selection we could find a Bell violations in the case of 1 photon loss on both sides, with $N=M-1$ (unanimity voting), starting at $M=5$. For details on this post-selection, see ref \cite{brunner08}.

\section{Indistinguishable photons}\label{indist}

In the first part of this work we showed how, using multiple independent pairs together with independent global measurement on all the photons produced, one could find a substantial CH violation, even in the presence of lots of pairs. But how good is this compared to a source producing the $2M$ photons altogether? For the sake of comparison we now consider a specific example, commonly produced in many labs. By the same occasion it will uncover some aspects of the relation between entanglement and non-locality.

The state we are discussing now can be written as $\boldsymbol{\rho}_M=\ket{\Phi_M}\bra{\Phi_M}$ with
\be \label{eq:phi} \ket{\Phi_M}=\frac1{M!\sqrt{M+1}}\left(a_0^\dag b_0^\dag + a_1^\dag b_1^\dag \right)^M\ket{0}\ee
where $a_0$, $a_1$ are orthogonal modes on Alice's side and $b_0$, $b_1$ orthogonal modes on Bob's side (for instance horizontal and vertical polarization modes). A way to produce this state is with a parametric down conversion (PDC) source, which gives a poissonian distribution of such states. The same global measurements as previously performed on $M$ photons can be realized here by just using the same setup as before : a polarizer followed by two photon counters on each side (same setup as represented in Figure \ref{fig:setup}, but with a different source).

Considering state \eqref{eq:phi} we make a similar analysis as previously, briefly reviewing the results of \cite{reid02,drummond83} for the amount of violation achievable, and presenting our own analysis for the resistance to noise.

\begin{figure}
\subfigure[]{\label{fig:indistingable}\includegraphics{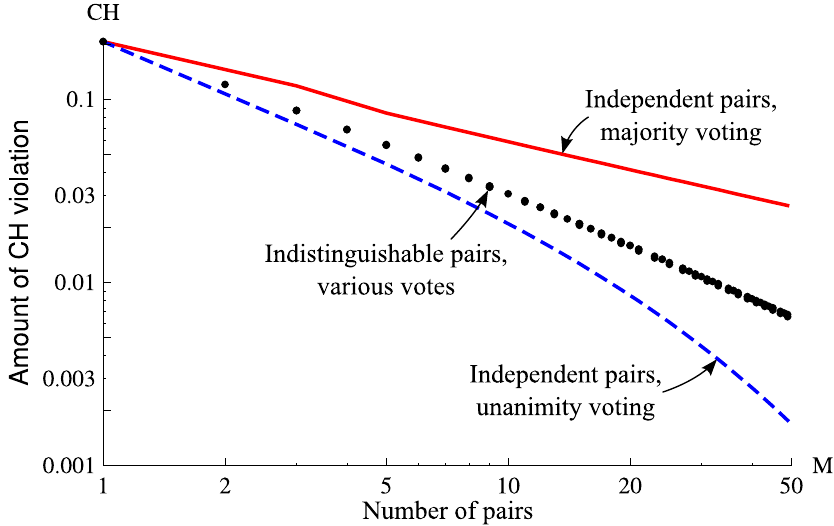}}
\subfigure[]{\label{fig:resistance}\includegraphics{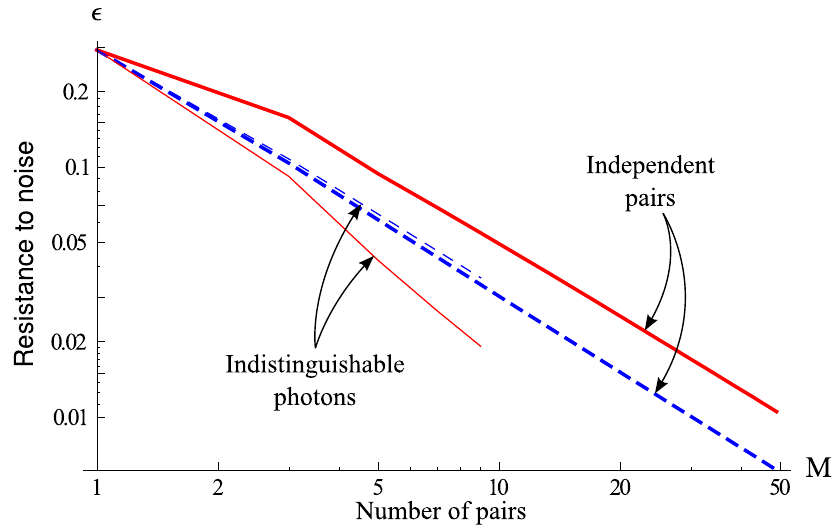}}

\caption{(Color online) Comparison between sources producing independent pairs or indistinguishable photons, using settings of the form \eqref{eq:settings}.\\
(a) Maximal CH violation achieved with a source of indistinguishable photons for various voting procedure (superposed black dots). Compared to the previously calculated violations for independent pairs (same curves as in figure \ref{fig:CH}), unanimity voting (lower blue dashed line) yields less violation, while majority voting (upper red line) yield the highest $CH$ values. Note that the maximal violation with indistinguishable photons almost doesn't depend on the voting procedure used.\\
(b) Maximal resistance to noise in the majority voting scenario (full red lines) and the unanimity scenario (dashed blue lines) for sources producing independent pairs (thick line, same curves as in figure \ref{fig:1mw}) or indistinguishable photons (thin line). The unanimous vote is more robust for indistinguishable photons, but majority voting on independently produced pairs yields the most persistent violation.}
\end{figure}

We computed the new probabilities entering the $CH$ expression for this specific state and, choosing various voting procedures, numerically optimized the settings according to $\alpha$ \eqref{eq:settings} in order to get the largest $CH$ violation. Surprisingly, for any number of photons $M$, all voting procedures yield approximately the same maximum violation of $CH$, decreasing as $M^{-1}$ (see figure \ref{fig:indistingable}). This is even more surprising as the settings needed for that are not the same for all voting methods.

Note that Reid et al. \cite{reid02} used another figure of merit : $S=\frac{CH+B}{B}$ (with $B=P_+^A(a_1)+P_+^B(b_1)$), which gives different results for the different voting strategies. Recalling the artefacts we already found in the amount of CH violation for poissonian sources, we choose to look now at an experimentally meaningful figure of merit, namely the resistance to noise.

\subsection{Noisy symmetric state}

Unlike for distinguishable photons, the effect of a noise map on a symmetric $M$-photons state does not affect each photon independently. We thus need here a more precise noise model. For the sake of simplicity, we put ourselves in an asymmetric setting, modeling the noise observed in the state measured by Alice and Bob as coming from the imperfection of the channel linking the source and Alice. Because the channel slightly deteriorates the systems passing through it but has no preferred basis, we model it by an average over all rotation axes $\vec n=(\sin\theta\cos\phi,\sin\theta\sin\phi, \cos\theta)$ in the Bloch sphere of rotations $U$ by an angle $2\beta$, with $\beta$ following a properly normalized gaussian distribution $p(\beta)=\frac2{\left(1-e^{-2\sigma^2}\right)\sqrt{2\pi}\sigma}e^{-\frac{\beta^2}{2\sigma^2}}$. For any representation $\vec\sigma=(\vec J_x,\vec J_y, \vec J_z)$ of SU(2) generators, the rotation operator is $U=\exp\left(-\beta \vec n\cdot \vec\sigma\right)$. The state after the noisy channel is thus given by~:
\ba
\boldsymbol{\rho}_{out} &=& \int_{SU(2)}p(\beta)(U\otimes\one)\boldsymbol{\rho}_{in}(U^\dag\otimes\one)[dU]\\
&=&\frac{1}{4\pi}\int_{-\infty}^{\infty}d\beta \int_{0}^{\pi}d\theta \int_{0}^{2\pi}d\phi \sin^2{\beta} \sin{\theta} p(\beta)\\
&& \times (U(\beta,\theta,\phi)\otimes\one)\boldsymbol{\rho}_{in}(U^\dag(\beta,\theta,\phi)\otimes\one)
\ea
where we have used the appropriate Haar measure of SU(2) in terms of the Euler angles~: $[dU]=\frac1{4\pi}\sin^2(\beta)\sin(\theta)d\beta d\theta d\phi$. We introduced the Haar measure here because it is the only measure which is invariant under group operations. It thus treats every rotation the same way, reflecting the fact that the noise has no preferred rotation axis.

This channel applied to a single pair state $\boldsymbol{\rho}_{in}=\ket{\Phi_1}\bra{\Phi_1}$ with $\ket{\Phi_1}=\frac1{\sqrt2}(a_0^\dag b_0^\dag + a_1^\dag b_1^\dag)\ket{0}$ produces a Werner state \eqref{eq:werner}, allowing one to make a correspondence between the usual noise model used in the previous part of this work, in terms of Werner states and this one~:
\be 3w = e^{-2\sigma^2}+e^{-4\sigma^2}+e^{-6\sigma^2}.\ee
This relation allows us to interpret the amount of white noise $\epsilon=1-w$ as being, to first order, the variance of the random rotation angle~:
\be\epsilon=4\sigma^2+O(\sigma^4).\ee

Applying this channel to the state $\ket{\Phi_M}$ for various $M$, and performing the majority and unanimity votes with settings in $\alpha$ \eqref{eq:settings}, we found that the unanimity procedure is more robust to noise than the majority vote, scaling like $\sim M^{-1}$ (see figure \ref{fig:resistance}).

\subsection{Particle losses}

To compare indistinguishable and independent pairs in the case of losses, we consider the case in which one particle is lost on each side, yielding a total number of detections $2(M-1)$. In terms of modes, the state measured after the loss of particles can be written
\be
\begin{split}
\boldsymbol{\rho}_{M-1}\sim &\ a_0b_0\ket{\Phi_M}\bra{\Phi_M}a_0^\dag b_0^\dag + a_0b_1\ket{\Phi_M}\bra{\Phi_M}a_0^\dag b_1^\dag\\
&+ a_1b_0\ket{\Phi_M}\bra{\Phi_M}a_1^\dag b_0^\dag + a_1b_1\ket{\Phi_M}\bra{\Phi_M}a_1^\dag b_1^\dag.
\end{split}
\ee
Using such a state, we could find a violation for sufficiently many pairs for $M\geq10$, starting majority voting. Thus, there is no need for additional post-selection here.

\section{Distinguishable vs. Indistinguishable pairs}\label{comparison}

In the last sections we examined how two different multiparticle bipartite states $\ket{\Psi_M}$ \eqref{eq:psi} and $\ket{\Phi_M}$ \eqref{eq:phi} could be used to show nonlocality using a natural setup producing binary outcomes. These two states are actually related~: if one were to produce the state of independent pairs $\ket{\Psi_M}$ with fundamentally indistinguishable photons on both Alice and Bob's sides, then the state created would be symmetric with respect to permutations between Alice's photons or Bob's ones, and we would actually have produced state $\ket{\Phi_M}$. This can be seen by projecting $\ket{\Psi_M}$ onto the corresponding symmetric subspaces~:
\be
\begin{split}\label{eq:sym}
\begin{split}
\ket{\Psi_M}=& \left(\frac{\ket{0}_A\ket{0}_B+\ket{1}_A\ket{1}_B}{\sqrt{2}}\right)^{\otimes M}\\
=& 2^{-\frac M2} \sum_{i=0}^M \sum_{\pi\in\Pi_i^M} \pi\ket{0}_A^{\otimes i}\ket{1}_A^{\otimes M-i}\otimes \pi\ket{0}_B^{\otimes i}\ket{1}_B^{\otimes M-i}\\
\end{split}\\
\xrightarrow{sym} 2^{-\frac M2} \sum_{i=0}^M \ket{i,M-i}_A\ket{i,M-i}_B \sim \ket{\Phi_M}
\end{split}
\ee
where $\Pi_i^M$ is the set of all $\binom{M}{i}$ possible arrangements of $i$ ``$0$'' and $M-i$ ``$1$'', and $\ket{i,M-i}_A=\frac{(a_0^\dag)^i(a_1^\dag)^{M-i}}{\sqrt{i!(M-i!)}}\ket{0}$ is the Fock state describing $i$ of Alice's $M$ photons in the ``$0$'' state and $M-i$ in the ``$1$'' state.

So the only difference between $\ket{\Psi_M}$ and $\ket{\Phi_M}$ is the distinguishability of the $M$ photons flying to Alice or Bob. But in the setup we considered (as described in figure \ref{fig:setup}), we didn't take advantage of the particular pairing between some of Alice's photons with some of Bob's ones. Because we applied global measurement, we could even suppose that all photons on Alice(Bob) side were mixed before reaching the beamsplitter. In other words we classically lost trace of the pairing between Alice's and Bob's photons. We are thus comparing a situation in which one explicitly chose not to distinguish between photons belonging to a given set, with another one for which these photons are intrinsically indistinguishable.

Let us now compare the entanglement present in both states. Eisert et al.\cite{eisert00} calculated the amount of entanglement present in the state of distinguishable particles $\ket{\Psi_M}$ after having forgotten the pairing of Alice's photons with Bob's ones. For $M$ even~:
\be E_d=E(\ket{\Psi_M}) = \sum_{j=0}^{M/2} \frac{(2j+1)^2}{2^M(M+1)}\binom{M+1}{M/2-j}\log_2(2j+1).\ee
Concerning the state of indistinguishable particles $\ket{\Phi_M}$, writing it in terms of modes as in equation \eqref{eq:sym} we see that its entanglement is given by
\be E_i=E(\ket{\Phi_M}) = \log_2(M+1)\ee
since it is a maximally entangled state of two systems of dimension $M$. Evaluating these two quantities, we find $E_i>E_d\ \forall M$, and more precisely $E_i/E_d\xrightarrow{M\rightarrow\infty} 2$. So more entanglement is present in the state where photons are quantumly indistinguishable, but a larger violation of CH can be observed using a natural setup if the photons are in principle distinguishable, but we choose not to make any difference between them. Looking at how resistant these violations are with respect to noise confirms this order. Only should it be noted that compared to particle losses, the indistinguishable case looks more resistant, since no additional post-selection was necessary to find a violation when both Alice and Bob lost a particle during the experiment.

This is in agreement with other results \cite{Methot07}, showing that entanglement and nonlocality are different measures.

\section{Experimental perspectives}\label{experiments}

In this Section we give a brief overview of experimental situation where our techniques might be applied.

As mentioned previously, the experiment of Ref. \cite{anderlini}
shows evidences for entanglement in ensembles of ultracold atoms
of $^{87}$Rb in an optical lattice. Entanglement between two
atomic levels is generated via a partial swap gate, an entangling
operation. In order to apply our techniques, the atoms of each
level should be addressed separately; that is Alice should hold
all atoms in the ground state, and Bob all atoms in the excited
state. Note that in this experiment the pairs are independent because 
they are located in different regions of the optical lattice.

Another experimental situation invoking Bose-Einstein condensates
where our techniques might be useful is superradiant scattering
\cite{superradiant}. It has been argued that this process
generates entanglement between the emitted photons and the atoms
of the condensates. In that case the particles would be
indistinguishable.

A third possibility is the experiment discussed in Ref.
\cite{electronhole}, which is a proposal for energy-time
entanglement of quasiparticles in a solid-state device. This
experiment is the adaptation of the Franson-type experiment 
\cite{franson89} with entangled electron-hole
pairs.

Finally it is also worth mentioning quantum optics. However it is
not clear that our techniques will turn out useful in this field,
since they require high detection efficiencies, a feature that
still lacks generally in optics. Still, sources producing independent entangled pairs, or indistinguishable photons via parametric down-conversion (PDC), are already well understood. Multi-photon entanglement using PDC sources was demonstrated in \cite{eisenberg04}. A careful analysis of post-selection might thus open the possibility to feasible experiments.

\section{Conclusion}

\begin{table}
\begin{center}
\includegraphics[width=0.48\textwidth]{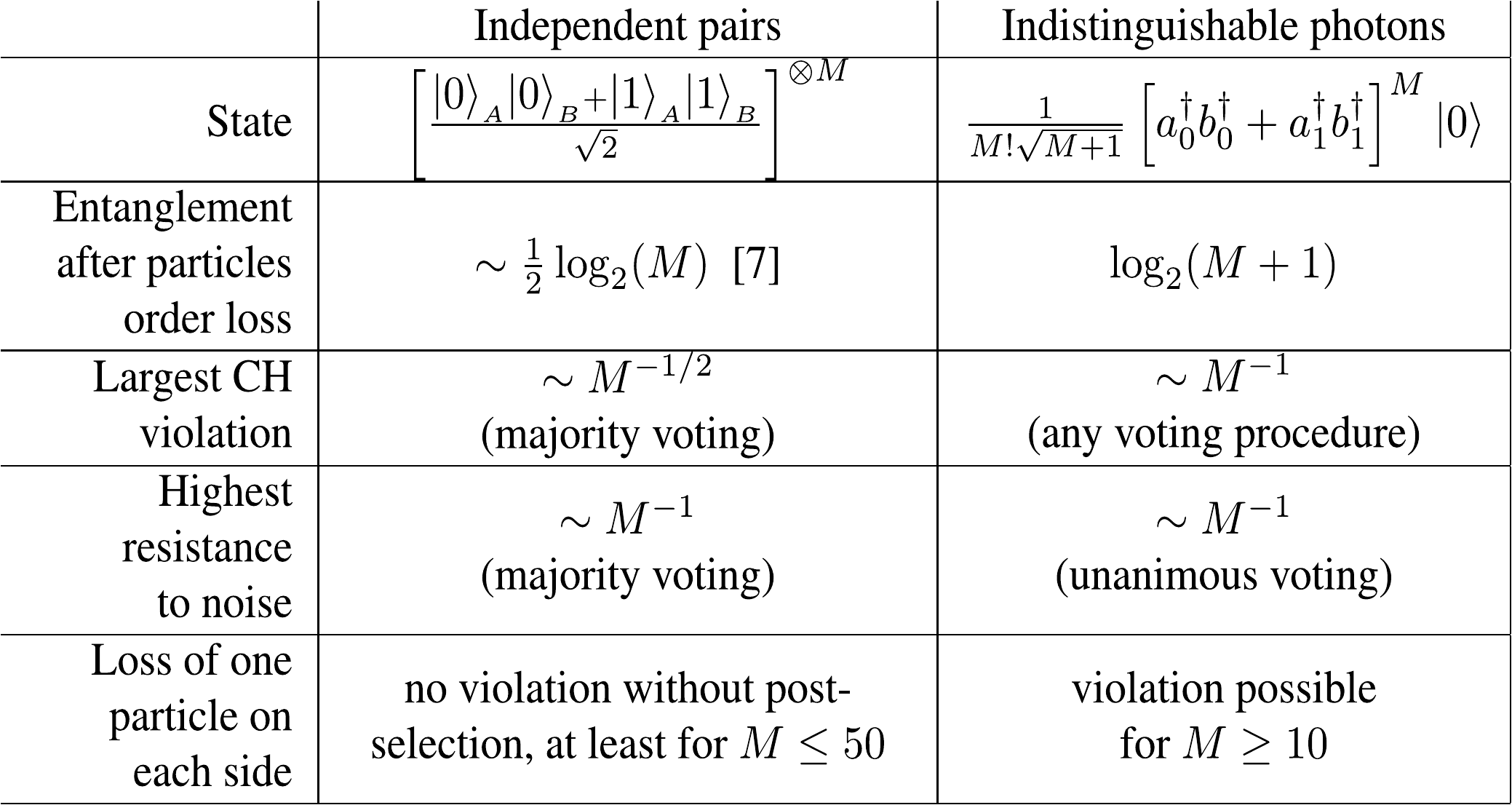}
\caption{Summary of the main results of this work.}
\label{table:summary}
\end{center}
\end{table}

We considered Bell experiments on multiple pairs of particles, where the two parties are not able to address each particle separately and thus call upon global measurements, projecting all of their incoming particles in the same basis. Votes were introduced as a natural way to produce binary outcomes from two detection numbers. This allowed us to test the CH inequality in the presence of both a source of $M$ independent pairs and of $M$ indistinguishable pairs, highlighting a violation of the CH inequality for any number of pairs $M$. Considering the resistance to noise of such violations, modeled as a noisy channel, we could provide an experimentally meaningful measure of nonlocality. The impact of losses was also evaluated for the two situations, showing that indistinguishable pairs are more robust against losses. More detailed results are summarized in table \ref{table:summary}. Finally, a comparison of the nonlocality observed for each source with the entanglement of their respective states provided another example of non-monotonicity between these two quantities.

\section{Acknowledgements}
We thank F.S. Cataliotti for pointing out the potential application of our results to superradiant scattering.

We acknowledge financial support from the EU project QAP (IST-FET FP6-015848) and the Swiss NCCR Quantum Photonics.


\end{document}